\documentclass[final,3p,times,twocolumn]{elsarticle}

\usepackage[T1]{fontenc}
\usepackage{booktabs}
\usepackage{amssymb}
\usepackage{lipsum}
\usepackage[dvipsnames]{xcolor}
\usepackage{amsmath}
\usepackage{soul}
\usepackage{graphicx}%
\usepackage{slashed}
\usepackage{graphicx}%
\usepackage{multirow}%
\usepackage{amsmath,amssymb,amsfonts}%
\usepackage{amsthm}%
\usepackage{mathrsfs}%

\usepackage{braket}
\usepackage{soul}
\usepackage{slashed}

\journal{Physics Letters B}

\begin{document}

\begin{frontmatter}

\title{The axial-vector  nucleon form factor in the meson dominance picture: \\ the role triangle singularities from a dispersive view}

\author[1]{Pablo S\'anchez Puertas}
\ead{pablosanchez@ugr.es}

\author[1]{Enrique Ruiz Arriola}
\ead{earriola@ugr.es}

\affiliation[1]{
organization={Departamento de F\'{\i}sica At\'{o}mica, Molecular y Nuclear and Instituto Carlos I de F\'{\i}sica Te\'{o}rica y Computacional},
addressline={Universidad de Granada},
postcode={E-18071},
city={Granada},
country={Spain}
}

\begin{abstract}

The nucleon axial form factor is analyzed in terms of a dispersion
relation comprising chiral perturbation theory (ChPT) and perturbative
QCD (pQCD) in the low- and high-momentum regimes respectively, which
implies a set of superconvergence sum rules for the $A \to N \bar N$
spectral function. In the timelike region below the $N \bar N$
production threshold we include the $a_1$ and $a_1'$ resonances, with
finite-width effects modeled through $\rho \pi$ and $\sigma \pi$
intermediate states.
Above the $N \bar N$ threshold, we model an infinite tower of radially
excited $1^{++}$ resonances with a non-integer power fall-off,
eventually matching pQCD. In this setup, we find that both the ChPT
and pQCD contributions are tiny. In turn, the dominant $\rho\pi$
triangle provides the leading deviations from axial meson
dominance. Our calculation is not a fit; it contains parameters with
{\it a priori} uncertainties estimates. The axial radius is predicted
to be $r_A^2 =(0.28-0.33)~\textrm{fm}^2 = ( 0.47-0.57~\textrm{fm})^2$
in agreement with recent lattice QCD results but inconsistent with the
MINER$\nu$A determination.  A simple semiempirical formula is provided
accounting for the main physical effects.
  
\end{abstract}


\end{frontmatter}

\bibliographystyle{elsarticle-num-names}

\section{Introduction}

The axial structure of the nucleon accounts for the coupling to
electroweak currents mediated by $W^\pm$ and $Z$ exchange (see
e.g.~\cite{LlewellynSmith:1971uhs,Gourdin:1974iq,Bernard:2001rs} for
reviews). On the experimental, side neutrino scattering and pion
electroproduction experiments have been pursued for many years with
varying degree of accuracy depending on nuclear and pion mass
corrections, respectively. The first high-statistics measurement of
$G_A$ has been made by the MINER$\nu$A collaboration directly on
nucleon targets and free from nuclear corrections or pion mass
extrapolations~\cite{MINERvA:2023avz}. More recently, first-principles
lattice QCD calculations have achieved realistic
physical conditions~\cite{Alexandrou:2020okk} (see e.g. lattice
summary \cite{Gupta:2024krt}). The story of discrepancies between
experiments and/or lattice QCD is rather long (see e.g.~\cite{MINERvA:2023avz}). 
Radiative corrections have also been
analyzed with an effect masked by other and larger
uncertainties~\cite{Tomalak:2026wsu}.  Recently, the MINER$\nu$A
collaboration~\cite{MINERvA:2025ygc} and lattice
QCD~\cite{Meyer:2026kdl} have firmly establish the discrepancy,
despite claims of high accuracy on each side. Actually, the comparison
rests on the z-expansion~\cite{Bhattacharya:2011ah}, which suitably
implements analyticity taking into account threshold singularities and
high energy behaviour. 

From a theoretical point of view, analyticity imposes useful and
general constraints on $G_A$. Dispersion relations for $G_A $ were
already proposed long ago \cite{Goldberger:1958vp} (see
\cite{barton1965introduction} for an early comprehensive
account). They exploit $t$-channel unitarity, corresponding to the $A \to
N \bar N$ process in the timelike region, which is characterized by the
corresponding spectral function 
with $I^G J^{PC}= 1^-1^{++}$ quantum numbers in the isospin limit.

In this 
paper we analyze  the axial nucleon form factor using
dispersion relations and hadronic input from $3 \pi$, $\pi
\sigma$, $\pi \rho $ and $a_1, a_1', \dots$ resonances in the $I^G
J^{PC}= 1^-1^{++}$ channel. We incorporate rigorous QCD features such
as analyticity, ChPT and pQCD and, more importantly, the
implementation of superconvergence sum rules, which are flagrantly
violated in many calculations. Our calculation improves on previous
meson dominance calculations~\cite{Masjuan:2012sk,Amaro:2015lga} both
in precision and accuracy, even after accounting of all known sources of
uncertainties, and without fitting the available $G_A$ data.

\section{Formalism}

We consider the matrix element of the axial current 
\begin{multline}
\bra{N_{s'}(p')} A_{\mu}^a \ket{N_s(p)} = \\ \bar{u}_{s'} (p')\left[ \gamma_{\mu} G_A(q^2) +q_{\mu}\frac{G_P(q^2)}{2m_N} \right] \gamma^5 \frac{\tau^a}{2} u_s (p), 
\end{multline} 
where $q_{\mu}=p'_{\mu}-p_{\mu}$ and $N=(p,n)$, with
$m_N=(m_n+m_p)/2$. The axial current is defined as $A_{\mu}^a =
\bar{q}\gamma_{\mu}\gamma^5\frac{\tau^a}{2}q $, with
$q^T=(u,d)$. The form factors $G_A(q^2)$ and $G_P(q^2)$ are known,
respectively, as the axial and induced pseudoscalar form factors.
The crossed process corresponds to
\begin{eqnarray}
\bra{ \bar N N}  A_{\mu}^a \ket{0} =  \bar u_{s'}(p')\left[ \gamma_{\mu} G_A(s) +P_{\mu}\frac{G_P(s)}{2m_N} \right] \gamma^5 \frac{\tau^a}{2} v_s(p),
\end{eqnarray}
with $s= (p'+p)^2 >0 $.

\subsection{Analytical properties}

The function $G_A(t)$ in the complex $t$-plane has relevant analytical properties, 
which we review next for completeness assuming isospin invariance.  
\begin{itemize}
\item Normalization,  $G_A(0) = g_A $ (the $n$  beta decay axial coupling constant, $n \to p + e^- + \bar \nu_e$)
\item It has a branch cut at $s= (3 m_\pi)^2$. Its 
discontinuity, computed from ChPT, behaves as~\cite{Bernard:1996cc,Kaiser:2019irl}
\begin{eqnarray}
  \operatorname{Disc}G_A(s) 
  = 2 i \operatorname{Im}G_A(s) \sim 2 i (\sqrt{s}-3 m_\pi)^2  \, ,
\label{eq:thr-GA}  
\end{eqnarray}
which is stronger than $P$-wave $\sim (\sqrt{s}-3 m_\pi)^3$ phase-space behaviour
\item $G_A(t)$ is real for spacelike momenta , $t=-Q^2 < 0$. This
  implies that, for complex $t$, $G_A(t)^* = G_A(t^*)$ (reflection
  principle).
\item It falls off as $(\alpha_s(-t)/ t)^2$ for $t \to - \infty$~\cite{Carlson:1985zu}. Analytic continuation
  to the timelike region, $s+ i \epsilon \equiv \lim_{\theta \to \pi}|t| e^{-i \theta} $, 
  then yields~\cite{RuizArriola:2025wyq,RuizArriola:2025omi} 
  \begin{eqnarray}\label{eq:pQCDIm}
  G_A (s+ i \epsilon) \sim \frac{(A_u+A_d)}{s^2} \left[\frac{4\pi/\beta_0}{\log (s/\Lambda^2)-i \pi}  \right]^2 
  \end{eqnarray}
where $\beta_0 = 11 N_c/3 - 2 N_f/3=9$ and $A_{u,d}$ are factors (see  Table 4 of Ref.~\cite{RuizArriola:2025omi} for a compilation)
\item It has poles at the PDG axial resonances $a_1, a_1', a_1'' , \dots$ in the second Riemann sheet  $ \sqrt{s} = m_{a_1} - i \Gamma_{a_1}/2 , m_{a_1'} - i \Gamma_{a_1'}/2 \dots $
\item It has a $N \bar N$ branch cut at $s \ge 4 M_N^2$ corresponding
  to elastic $N \bar N$ scattering in the $^3P_1$ channel.
\end{itemize}

\subsection{Re-Im dispersion relations}

With these properties we have the unsubtracted dispersion relation for
spacelike momenta
\begin{eqnarray}
G_A(-Q^2 ) = \frac1{\pi} \int_{(3 m_\pi)^2}^\infty \frac{\operatorname{Im}G_A(s)}{s+Q^2} ds \, . 
\end{eqnarray}
From the conditions
\begin{eqnarray}
 \lim_{Q^2 \to \infty} Q^2 G_A(-Q^2)= \lim_{Q^4 \to \infty} Q^4
 G_A(-Q^2)= 0 
\end{eqnarray}
and the normalization, we have the following sum rules
\begin{eqnarray}
  g_A &=&   \frac1{\pi} \int_{(3 m_\pi)^2}^\infty \frac{\operatorname{Im}G_A(s)}{s} \, ds \, , \\
  0   &=&   \frac1{\pi} \int_{(3 m_\pi)^2}^\infty \operatorname{Im}G_A(s) \, ds \, , \\
  0   &=&   \frac1{\pi} \int_{(3 m_\pi)^2}^\infty \operatorname{Im}G_A(s) s \, ds \, . 
\end{eqnarray}

The last two sum rules imply separately that the corresponding
spectral function is {\it not} positive definite. However, we show
next that they jointly imply that there are at least {\it two
  zeros}. To show this, let us assume that $s_1$ is the only zero, so
that the spectral function has opposite signs in the regions
$s_0 \le s \le s_1 $ and $s \ge s_1$. Then, from the last sum rules,
we obtain an equality of mean values
\begin{eqnarray}
\langle s \rangle_{\rm left} \equiv \frac{\int_{s_0}^{s_1} \operatorname{Im}G_A(s) s \, ds}{\int_{s_0}^{s_1} \operatorname{Im}G_A(s) ds} 
      = \frac{\int_{s_1}^\infty \operatorname{Im}G_A(s) s\,  ds}{\int_{s_1}^\infty \operatorname{Im}G_A(s)\,  ds} \equiv \langle s \rangle_{\rm right}
\end{eqnarray}
which is a contradiction since $ s_0 < \langle s \rangle_{\rm left} <
s_1$ whereas $ s_1 < \langle s \rangle_{\rm right} $. Thus, the
assumption of only one zero is incorrect.

\subsection{Phase-modulus dispersion relations}

An advantageous way of imposing the conditions is by using the
modulus-phase representation, which in the timelike region 
reads $G_A(s)= |G_A(s)|e^{i \delta_A(s)}$. Therefore,
\begin{equation}
\frac{G_A(s+i\epsilon)}{G_A(s-i \epsilon)}= e^{2 i \delta_A(s+i\epsilon)}  , \quad
\tan \delta_A(s) = \frac{\operatorname{Im}G_A(s)}{{\rm Re} G_A(s)} \, . 
\end{equation}  
The threshold behaviour follows from Eq.~\eqref{eq:thr-GA},
\begin{equation}
\delta_A(s)
\sim (\sqrt{s}-3 m_\pi)^2 \, ,
\end{equation}
whereas its asymptotic behaviour is obtained from the argument 
principle on the first Riemann sheet, following the construction 
of Ref.~\cite{RuizArriola:2024gwb}. To this end, we consider a closed
positive contour of radius $|s|=\Lambda^2 \to \infty $ and argument $
-\pi < \theta < \pi $ encircling the branch cut $s > (3 m_\pi)^2 $.
For an analytical function with $N$ zeros and $P$ poles inside 
the contour, one has 
\begin{eqnarray}
  N-P &=& \int_{C_I} \frac{dz}{2\pi i} \frac{G_A'(z)}{G_A(z)} =  + 2 i \frac{\delta_A( \Lambda^2) - \delta_A(s_0))}{2\pi i} 
  \nonumber \\
  &+&   \int_{-\pi}^{\pi} z \frac{d \theta}{2\pi } \frac{d}{dz} \log G_A(z) |_{z= \Lambda^2 e^{i \theta}} \, .
\end{eqnarray}
Evaluating the integral and using $\delta(s_0)=0$ one finds for the phase
at large timelike $s \to \infty$ 
\begin{eqnarray}
  \delta_A(s+ i \epsilon)/\pi = 2 \left[ 1 + \frac1{2\log(s/\Lambda_{\rm QCD}^2)} \right]+ N-P
\end{eqnarray}
Using $\operatorname{Im}G_A(s)= \sin \delta_A(s) |G_A(s)|$, 
the first two zeros are located at $\pi= \delta_A(s_1)$ and 
$2\pi= \delta_A(s_2)$.
While there
are no poles ($P=0$), the number of zeros in the complex plane is unknown
$(N \ge 0)$, so pQCD implies that the spectral function has at least
two zeros. Phenomenologically, we interpret the Breit-Wigner resonance
axial meson masses $\delta_A (m_{a_1}^2) = \pi/2$ and $\delta_A
(m_{a_1'}^2) = 3\pi/2$ and widths $ \Gamma_{a_1} = 1/( m_{a_1} \delta_A'(m_{a_1}^2) ) $ as the local extrema nested between the zeros, $s_0 \le m_{a_1}^2 \le s_1 \le
m_{a_1'}^2 \le s_2 $ 

Actually, the argument principle in the second Riemann
sheet leads to $ 2 +
N_{\rm II}- P_{\rm II}= -(2 + N_{\rm I}- P_{\rm I}) $, so that $P_{\rm
  II}= 4+ N_{\rm I}+N_{\rm II} + P_{\rm I} \ge 4$, indicating the 
presence of {\it least} 4 poles in the second Riemann sheet.
Since $G_A^{\rm II}(z^*) = G_A^{\rm I}(z) = G_A^{\rm I}(z^*)^* = G_A^{\rm II}(z)^* $, 
these poles occur in complex-conjugated pairs, that we identify with the $a_1$
and $a_1'$ resonances.

Since, at present, there is no evidence of any bound $N \bar N$ state 
($P_I=0$) nor a zero of the form factor ($N_I=0$), we may use the 
once-subtracted Omn{\`e}s representation
\begin{equation}
\frac{G_A(t)}{g_A}= \Omega_A(t) =  \exp \left[\frac{t}{\pi} \int_{s_0}^\infty
  \frac{\delta_A(s)}{s(s-t-i \epsilon)}\right] \, .
\end{equation}

\section{Anatomy of the spectral functions}

The spectral function is determined from the $N \bar N$ processes with
$1^{++}$ quantum numbers. In the isospin limit, and due to G-parity,
these are the processes with an odd number of pions, which are however
unphysical (not accessible experimentally) below the $N\bar N$ 
threshold. Following previous studies, we divide our study into
several regions. While this separation is well defined experimentally,
it is not obvious how to deal theoretically with possible overlapping
issues, particularly the resonance backgrounds, in order not to double
count. While some modeling, which we postpone to Sect.~\ref{sec:numres}, is
unavoidable, the sum rules provides some overall control.

\subsection{Model independent features }

We establish first that the extreme values corresponding to the
threshold region and the asymptotic region can be described by ChPT
and pQCD respectively, but ultimately provide tiny contributions to
the axial form factor $G_A(t)$. This is in agreement with our previous
finding on the pseudoscalar form factor~\cite{RuizArriola:2025omi}.

\subsubsection{Threshold region: chiral perturbation theory}

The threshold behaviour is constrained by chiral symmetry, as first 
proposed by Pagels~\cite{Pagels:1972xx} and revisited using 
HBChPT~\cite{Bernard:1996cc} and relativistic formulations with and 
without $\Delta$ degrees of freedom~\cite{Kaiser:2019irl}. For our 
purposes, the simple semiempirical 
formula 
\begin{eqnarray}
  \operatorname{Im}G_A^{\rm ChPT}(s) \approx \frac{(s-9m_\pi^2)^2 g_A}{9216 \pi ^3 f_{\pi }^4}\left(\left(1+\frac{64 \pi ^2}{35}\right) g_A^2 -1\right)
\end{eqnarray}
is good enough; it reproduces the chiral limit~\cite{Bernard:1996cc},
incorporates the proper threshold behaviour and accounts for $90\%$
of the LO non-relativistic result~\cite{Bernard:1996cc} in the credible
range $ 3m_\pi \le \sqrt{s} \le 6 m_\pi$.\footnote{The relativistic
  calculation with $N-\Delta$ degrees of freedom~\cite{Kaiser:2019irl}
  provides a relative correction of $30\%$.} Importantly, all these
variants find a \textit{positive} contribution to the spectral
function.  Numerically,
\begin{eqnarray}
  G_A^{\rm th}(-Q^2) =  \int_{9 m_\pi^2}^{\Lambda_\chi^2} \frac{ds}{\pi} \frac{\operatorname{Im}G_A^{\rm ChPT}(s)}{s+Q^2} \le  G_A^{\rm th}(0) 
\end{eqnarray}
provides a negligible contribution for $\Lambda_\chi \le 6 m_\pi$, 
well below $1\%$ in the $0 \le Q^2 \le 2~{\rm GeV}^2$ region.

\subsubsection{Asymptotic region and pQCD}

The pQCD condition is a genuine feature of QCD, as already shown in 
Eq.~\eqref{eq:pQCDIm}. The corresponding contribution for 
$s > \Lambda_{\rm pQCD}^2$ reads
\begin{eqnarray}
  G_A^{\rm pQCD}(-Q^2) =  \int_{\Lambda_{\rm pQCD}^2}^{\infty} \frac{ds}{\pi} \frac{\operatorname{Im}G_A(s)}{s+Q^2} \, .
\end{eqnarray}
There is a huge disparity of predictions in the numerical value of
the nucleon amplitude (see Table 4 of Ref.~\cite{RuizArriola:2025omi}), and even the
sign seems disputed. Still, even for the largest values, the dispersive
contribution is less than $1\%$ at the  unreasonable lowest 
pQCD limit of $N\bar N$ threshold, $\Lambda_{\rm pQCD}^2=4 m_N^2$.

\subsection{Intermediate-energy region: resonances and second Riemann sheet singularities}

In the intermediate-energy regime, the axial-vector resonances with
$I^G J^{PC}=1^-1^{++}$ quantum numbers are expected to dominate the
spectral function over the $3\pi$ nonresonant continuum.
Nevertheless, finite-width effects must in principle be taken into
account.  These comprise both the universal right-hand cut
contributions associated with rescattering effects and, more
importantly, process-dependent left-hand-cut (triangle) singularities.
Since a full three-body unitarization is a daunting enterprise, in the
following we adopt a quasi-two-body $R\pi$ description motivated by
the dominant $a_1\to R\pi$ decays, where $R=\rho$
dominates~\cite{ParticleDataGroup:2024cfk}.

\subsubsection{Resonances and right-hand cuts}\label{sec:RHC}

We remind that the representation of resonances, except for the
rigorous conditions of producing poles in the second Riemann sheet
and the appropriate threshold behaviour,
leaves still some ambiguity in the timelike region. 
Only in the narrow-resonance approximation we have a
universal spectral function,  
\begin{eqnarray}
\operatorname{Im}G_A(s) = Z_{a_1} \delta(s-m_{a_1}^2) + Z_{a_1'} \delta(s-m_{a_1'}^2) \, ,
\end{eqnarray}
for $m_\pi+m_R \le \sqrt{s} \le 2m_N $. Relaxing this condition to
account for finite width effects, analyticity and threshold behaviour,
we may take instead
\begin{eqnarray}
  \operatorname{Im}G_A^{\rm Res}(s) = Z_{a_1} \operatorname{Im}D_{a_1} (s) + Z_{a_1'} \operatorname{Im}D_{a_1'} (s) \, ,
\end{eqnarray}
where $D_a (s) = 1/(s-m_a^2 - \Pi_a (s))$ with $\Pi_a(s)$
real-analytic and $\operatorname{Im}\Pi_a(s) = m_a \Gamma_a(s)$ for
$s> (m_\pi + m_R)^2$.  The modeling of $\Pi_a(s)$ in terms of
quasi-two-body $R\pi$ intermediate states is postponed to Sect.~\ref{sec:resLS},
where the details of the unitarization procedure are discussed.

\subsubsection{Triangles and left-hand cuts}\label{sec:LHC}

\begin{figure*}[t]\centering
  \includegraphics[width=0.8\textwidth]{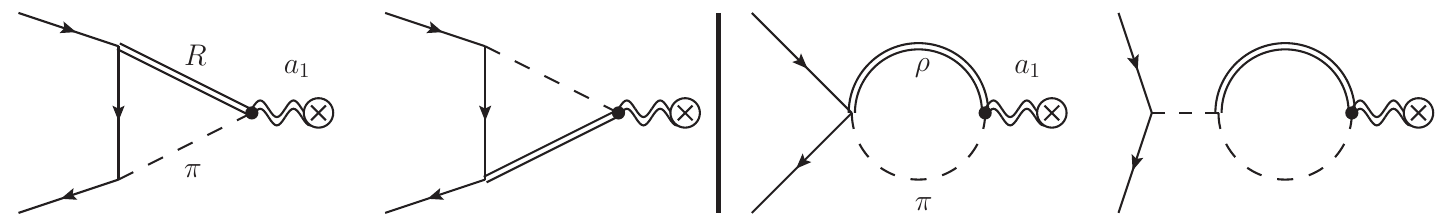}
  \caption{Model of the LHC based on $R\pi$ contributions
    ($R=\rho,\pi$).  The first two diagrams contribute to $\sigma$ and
    the Pauli $\rho N\bar{N}$ coupling.  The last two diagrams are
    necessary for the Dirac $\rho N\bar{N}$ coupling to ensure
    transversity, yet the last one does not contribute to $G_A$ (see
    details in the text).\label{fig:RpiLHC}}
\end{figure*}

An important process-dependent source of finite-width effects arises 
from left-hand-cut contributions. Within a quasi-two body $R\pi$ 
approximation, these are generated by the triangle diagrams in 
Fig.~\ref{fig:RpiLHC}. Factoring out the $a_1$ lineshape, to be 
reinstated after the unitarization procedure is discussed in 
Sects.~\ref{sec:Unitarization} and \ref{sec:numres}, the corresponding 
triangle contribution in terms of standard loop functions reads
   \begin{eqnarray}
      \operatorname{Im}G_A^{\rho \pi {\displaystyle\triangleleft}} (s)&=&      
      \frac{g_A g_{a\rho\pi}F_a}{8\pi^2 m_{a} F_\pi} \Big[ f_{\rho NN} \operatorname{Im} C_{00}^\rho  \nonumber \\\qquad  & & + (g_{\rho NN} + f_{\rho NN}) 2m_N^2  \operatorname{Im} C_1^\rho \Big] \, ,
      \\  \operatorname{Im}G_A^{\sigma \pi {\displaystyle\triangleleft}} (s)&=&   \frac{g_A g_{a\sigma\pi} g_{\sigma NN} m_N}{2\pi^2 m_{a}} \operatorname{Im}C_{00}^\sigma  \, ,
   \end{eqnarray}
where, for $s_{\textrm{th}} = (m_\pi+m_R)^2 < s< 4m_N^2 $,  
   \begin{eqnarray}
      \operatorname{Im}C_1^R &=& 
        \frac{2\pi}{s(s-4m_N^2)}\Big[\frac{s -m_\pi^2 -m_R^2}{\sigma_N} T_R -\lambda^{1/2}_{\pi R}\Big] \\ 
      \operatorname{Im}C_{00}^R &=& 
      \frac{\pi}{4s(s-4m_N^2)}\Big[(s -m_R^2 -m_\pi^2)\lambda_{\pi R}^{1/2}  \nonumber \\
    &&    -\frac{4(m_N^2\lambda_{\pi R} +s m_\pi^2 m_R^2)}{s\sigma_N} T_R \Big]  \, .
   \end{eqnarray}
   \begin{eqnarray}
      \lambda_{\pi R} &=& [s^2 +(m_\pi^2 -m_R^2)^2 -2s(m_\pi^2 +m_R^2)] , \\ \sigma_N &=& \sqrt{4m_N^2/s -1} ,\\
         T_R &=&  \tan^{-1} \left( \frac{\lambda^{1/2}_{\pi R}\sigma_N}{s -m_\pi^2 -m_R^2} \right) .
   \end{eqnarray}
The real parts can be obtained from Cauchy's theorem.\footnote{The $C_{00}$ 
function is UV-divergent and, whenever is needed, we subtract it at zero. 
The undetermined subtraction constant absorbs point-like coupling effects. }
In the previous expressions, $g_{\rho NN}$ and $f_{\rho NN}$ are the 
Dirac and Pauli $\rho N\bar N$ couplings~\cite{Masjuan:2012sk}, while 
$g_{\sigma NN}$ is the $\sigma N\bar N$ one~\cite{CalleCordon:2009pit},
$F_{\rho,a}$ are the decay constants, and $g_{a R\pi}$ the $R\pi$ 
couplings to the $a_1$. For further details of vertices, definitions, 
and numerical inputs, we refer to \ref{app:LoopAndInputs}.
Note that, besides the discontinuity at the normal threshold 
$(\sqrt{s}= m_\pi +m_R)$, the imaginary part features 
a log-singularity in the second Riemann sheet, corresponding to the 
argument of $\arctan$ being $\pm i$, which is located below the normal 
threshold and relates to the left-hand cut.
In the limit of heavy nucleon it becomes\begin{eqnarray}
\sqrt{s_{\pi R}}|_{II}= (m_\pi + m_R) \left[ 1 - \frac{m_\pi m_R}{8 m_N^2} +
  \dots \right]
\end{eqnarray}  
This nearby singularity does affect strongly the $\pi R$ threshold 
behaviour, which radius of convergence is indeed small;  
numerically, $R_{\pi \rho}=1-\sqrt{s_{\pi \rho}}/(m_\pi+m_\rho)|_{II}=0.017$ 
and $R_{\pi \sigma}= 1-\sqrt{s_{\pi \sigma}}/(m_\pi+m_\sigma)|_{II}=0.015$.

\subsection{Overlapping resonance region above $N \bar N$: radial Regge trajectories}

The physics below the $N \bar N$ threshold is insufficient to fulfill
the superconvergence sum rules and, as discussed before, pQCD
extrapolated to $N\bar N$ is insufficient as well. While a finite
number of additional resonances could fulfill them, they would yield
an erroneous overdamped $t^{-3}$ behaviour.  This can be avoided if
considering an infinite tower of resonances in the context of radial
Regge trajectories~\cite{Masjuan:2012gc}, that find theoretical
support in quark models, AdS/CFT-inspired models, or quark-hadron
duality.
In the following, we use the scheme proposed in Refs.~\cite{RuizArriola:2025omi,RuizArriola:2026qiw} 
to model the infinite tower of $1^-1^{++}$ states as 
radial excitations of mesons for $2 m_N \le \sqrt{s} \le
  \Lambda_{\rm pQCD}$ as 
\begin{eqnarray}
  \operatorname{Im}G_A(s) &=& {\rm Im} {G_A (4m_N^2)} \left( \frac{4m_N^2}{s}
  \right)^{2+2 \epsilon} \, . 
\end{eqnarray}
As already mentioned, the numerical value of the asymptotic nucleon
wave function is quite uncertain, so, following \cite{RuizArriola:2025omi,RuizArriola:2026qiw}, 
we match the log-derivative at the pQCD scale, yielding
\begin{eqnarray}
\epsilon &=& \frac1{\log (\Lambda_{\rm pQCD}^2/\Lambda_{\rm QCD}^2)} \, , 
\end{eqnarray}  
and $\epsilon=0.1-0.2$ for the range $2 m_N \le \Lambda_{\rm pQCD} \le 10^6~{\rm GeV} $ 
(see e.g. Ref.~\cite{RuizArriola:2026qiw}).

\section{Unitarization and the Omn{\`e}s representation\label{sec:Unitarization}}

In principle, one could solve the dispersion relations above. 
However, the $R\pi$ resonance contributions 
in Sect.~\ref{sec:RHC} are intertwined, through unitarization, 
with the triangles discussed in Sect.~\ref{sec:LHC}.
For a point-like $R\pi\to N\bar{N}$ coupling, unitarization 
simply leads to an Omn{\`e}s function. 
In the present case, we instead regard the triangle as the 
irreducible $R\pi\to N\bar N$ transition, while the point-like 
coupling is absorbed in the subtraction constant.
In this way, the triangle model incorporates the left-hand cut 
in the corresponding inhomogeneous Omn{\`e}s problem. 
Since the triangle model is intended only for the low-energy region, 
we do not extrapolate it beyond its range of validity. Instead, 
we adopt a simple phase--modulus representation, incorporating 
the triangle phase along with the different contributions above as 
\begin{eqnarray}
  \delta_A (s) &=& \delta_{\rm ChpT}(s) + \delta_{\pi \rho 
  {\displaystyle \triangleleft}} (s)+\delta_{\pi \sigma  {\displaystyle
    \triangleleft}} (s) 
+  \delta_{a_1}(s)\nonumber \\ 
  & & + \delta_{a_1'}(s) + \delta_{\rm Regge} (s) + \delta_{\rm pQCD}(s) \, . 
\end{eqnarray}
The advantage of this additive representation is
that the different effects are multiplicative,
\begin{eqnarray}
  G_A(-t) = g_A \Omega_{\rm ChPT} \Omega_{ \pi \rho {\displaystyle  \triangleleft}}
  \Omega_{ \pi \sigma {\displaystyle  \triangleleft}}
 \Omega_{a_1} \Omega_{a_1'} \Omega_{\rm Regge} \Omega_{\rm pQCD}   \, ,
\end{eqnarray}
where every single factor is the corresponding Omn{\`e}s function.
This representation neatly displays 
the relevant analytical properties, including the log-singularities and 
the poles in the second Riemann sheet, and naturally incorporates the  
normalization and asymptotic sum rules. 

From our estimates earlier we disregard for the rest of the paper the
tiny contributions stemming from the threshold region and the pQCD
region, where the phases would become
\begin{eqnarray}
\frac{\delta_{\rm ChPT} (s)}{\pi} & \approx& \frac{32(s-9m_\pi^2)^2}{9 (16 \pi f_{\pi })^4}\left(\left(1+\frac{64 \pi ^2}{35}\right) g_A^2 -1\right) \\ 
\frac{\delta_{\rm pQCD} (s)}{\pi} & \approx & 2  \left[ 1 + \frac{1}{\log(s/\Lambda_{\rm QCD}^2)}  \right]  
\end{eqnarray}
respectively. They produce produce effects well below the accuracy of the intermediate energy region.

\section{Numerical results}
\label{sec:numres}

\begin{figure}[ttt]
\begin{center}
  \includegraphics[angle=0,width=0.42 \textwidth]{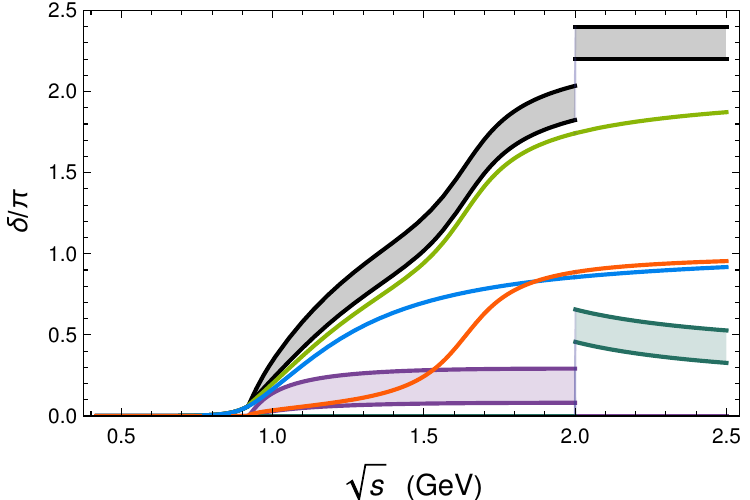} \\ 
 \includegraphics[angle=0,width=0.42 \textwidth]{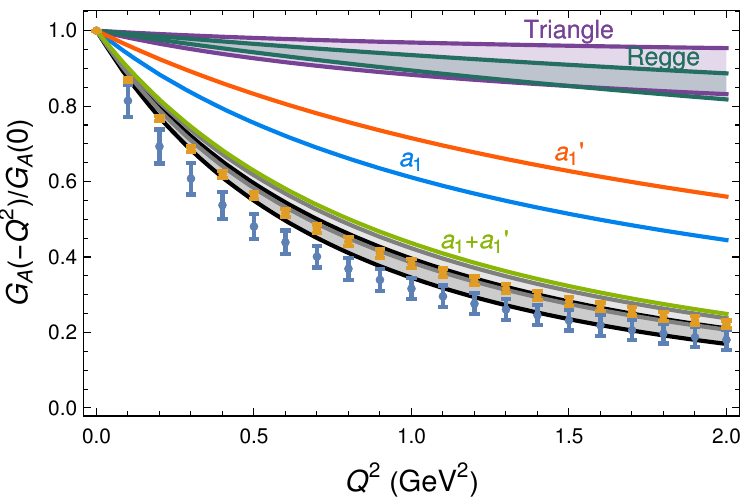} 
\end{center}
\caption{Upper panel: Spectral axial phases with the several
  contributions displayed separately. Lower panel: Same as before for
  the Omn{\`e}s reconstructed form factor in the spacelike region. We also show
the results from the z-expansion summary for lattice QCD and MINER$\nu$A 
experiments discussed in Refs~\cite{MINERvA:2025ygc,Meyer:2026kdl}.}
\label{fig:GA-all}
\end{figure}

\subsection{$\pi \rho$ and $\pi \sigma $ Triangles}

Regarding the triangle, we take 
\begin{eqnarray}\label{eq:TRphase}
\delta_{\pi R {\triangleleft}} = \tan^{-1} \left[\frac{ \operatorname{Im}G_A^{\pi R {\triangleleft}} (s)}
{ {\rm Re} G_A^{\pi R {\triangleleft}} (s)}\right] \approx \tan^{-1} \left[\frac{ \operatorname{Im}G_A^{\pi R {\triangleleft}} (s)}
{ g_A(1+ c_{\pi R} ) }\right]
\end{eqnarray} 
where the constant $c_{\pi R}$ accounts for a suitable combination of
the relevant couplings encompassing jointly the indiscernible contact
plus subtraction constant in the dispersion relation. This
simplification incorporates the smallness of the phase close to
threshold and is also in the spirit of the higher energy suppression
inherent to the Omn{\`e}s function. Note that, for $c_{\pi R} \gg 1 $, the
triangle phase vanishes. On the opposite, for $c_{\pi R}=-1$ we get $
\delta_{\pi R {\triangleleft}}= \pi/2 $, which acts as an
upper bound.  An estimate of the systematic uncertainty associated to
this scheme is further provided by the validity of the approximation
$\tan^{-1} x \approx x $, which holds for $x \lesssim 1$ and
corresponds to the range $(1+ c_{\pi R}) \gtrsim 2 $. This is a
necessary price we pay for the simplicity of the scheme.~\footnote{We
  have checked that, retaining the full phase in \eqref{eq:TRphase}, 
  one obtains similar results. Furthermore, we have analyzed the 
  canonical scheme base on the Re-Im dispersion
  relation. While it seems conceptually simpler, it is actually
  numerically more involved and actually not very different if
  uncertainties are placed in the calculation.}
In Fig.~\ref{fig:GA-all}, we show as a purple band 
the phase and modulus in the spacelike region for the range 
$2 \leq (1+ c_{\pi R}) \leq 10 $.

The approximations made allows, in the spacelike region, for a handy 
semiempirical representation which takes
into account the main effects of the $\pi \rho $ triangle and its
uncertainties. This is obtained from the Omn{\`e}s function assuming an average
phase (according to the mean value theorem)
$\overline{\delta}_{\triangleleft}/\pi$
\begin{eqnarray}
\Omega_{\pi R {\triangleleft}} \approx \left[ \frac{1+Q^2/4
    m_N^2}{1+Q^2/(m_\pi+m_R)^2}
  \right]^{\overline{\delta}_{\triangleleft}/\pi}
\end{eqnarray}
where $\overline{\delta}_{\triangleleft} \in
[0.2,0.72] $ for $\pi \rho$ and $\overline{\delta}_{\triangleleft} \in
[0.02,0.08] $ for $\pi \sigma$ nicely reproduce the spacelike result.

\subsection{Resonances below $N \bar N$\label{sec:resLS}}

Since the $a_1(1230)$ decays mostly to an $S$-wave $\rho\pi$
state~\cite{ParticleDataGroup:2024cfk}, we model its self-energy
through the corresponding standard 2-point loop function.  This is, to
a first approximation, $\Pi_{a_1}(s) = (g^2/2)
B_0(s,m_\pi^2,m_\rho^2)$, which is subtracted at $s=0$.  Further, to
implement the $\rho\to\pi\pi$ finite-width effects, we convolute it
with the $\rho$ spectral function, along similar lines to
Ref.~\cite{JPAC:2018zwp}, in order to obtain the final model for
$\Pi_{a_1}(s)$. We choose the parameters $m_{a_1}, g^2$ to obtain a
result that reproduces well those in Ref.~\cite{JPAC:2018zwp}. The
phase obtained this way, $\delta_{a_1}(s)$, is the one employed in the
Omn{\`e}s reconstruction.

To facilitate the comparison with the axial meson dominance, we
further extend the Omn{\`e}s integral range from $4m_N^2 \to \infty$,
accounting for this later on in the Regge regime appropriately (see
comments below).  The results for the phase and $\Omega_{a_1}$ in the
spacelike region are shown in blue in Fig.~\ref{fig:GA-all}.

By contrast to the $a_1(1260)$, which is well-known from $\tau\to3\pi$
decays, the experimental information for the $a_1(1640)$ is scarce,
and even its main decay channels remain unknown.  For simplicity, we
assume a $\rho\pi$ dominant decay channel,\footnote{The $a_1(1640)$ 
does not appear prominently in the $3\pi$ spectra~\cite{Rabusov:2023tna} 
and the $\rho'\pi$ channel could be important~\cite{Chen:2015iqa}. 
Given that $m_{\rho'} +m_\pi \sim m_{a_1'}$ and the large 
$\rho(1450)$ width, an appropriate description of its lineshape would 
also be required. Finally, the connection to the parameters in the 
PDG for each scenario is uncertain. For simplicity, we adopt the 
$\rho\pi$ decay channel.} 
adopting $\Pi_{a'_1}(s) = m_{a'_1}\Gamma_{a'_1}
B_0(s,m_\pi^2,m_\rho^2)/B_0(m_{a'_1}^2,m_\pi^2,m_\rho^2)$.  The
corresponding phase is employed in the Omn{\`e}s solution and, again,
we integrate up to infinity as in the $a_1(1260)$ case. The results
for its phase and modulus are shown in orange in
Fig.~\ref{fig:GA-all}.  Further experimental information would be
valuable for any dispersive reconstruction.

\subsection{Radial Regge resonances above $N \bar N$}

From our discussion above, for $ s \ge 4 m_N^2$, we have
$\delta_{\rm Regge}(s) = 2 \pi \epsilon $.  For large $\Lambda_{\rm pQCD} \gg 4 m_N$
we may extend the integral to infinity to  get 
\begin{eqnarray}
\Omega_{\rm Regge} = \frac1{(1+Q^2/4 m_N^2)^{2 \epsilon}} \, .
\end{eqnarray}
However, since $\delta_{a_1,a'_1}$ phases are integrated up to
infinity, we modify the corresponding Regge phase such that
$\delta_{\textrm{Regge}} + \delta_{a_1} +\delta_{a'_1} =
2\pi\epsilon$.  The corresponding phase is shown as a dark-green band
together with the corresponding Omn{\`e}s function in
Fig.~\ref{fig:GA-all}.

\subsection{Final results}

The results for each of the different pieces are shown in
Fig.~\ref{fig:GA-all}.  Ignoring triangle and Regge effects, the pure
meson-dominance model with energy-dependent widths is represented by the
light-green lines, which is far from experimental data and lattice QCD
results. Incorporating the triangle effects, the result is shifted to
the light-grey band and brings the results in reasonable agreement with
lattice QCD. Including Regge-physics effects, we obtain the final
dark-gray band within black lines, which still cannot reconcile
experimental data.  We emphasize that, whereas the heavy nature of
Regge physics is well reproduced through a linear effect, the triangle
one produces the require bending at low-energies required to better
reproduce lattice QCD results. Moreover, the axial radius becomes
\begin{eqnarray}
  \langle r^2 \rangle_A = \frac{6}{\pi} \int_{(3m_\pi)^2}^\infty \frac{\delta_A(s)}{s^2}
\end{eqnarray}
which is purely additive for each contribution in our
scheme. Numerically we find, in fm$^2$ units,
\begin{align}
  \langle r_A^2 \rangle & = 0.152 +0.096 +(0.014-0.053) +(0.017-0.029) \nonumber \\ 
                       & =(0.28-0.33)~\textrm{fm}^2 \, ,
\end{align} 
where each individual contribution arises from the $a_1$, $a'_1$,
triangle, and Regge, respectively.  The radius extraction in
experiment or lattice QCD analysis has a strong dependence on the
chosen interpolation function. For a compilation of recent results,
see Ref.~\cite{Alvarado:2026kny}. Our result is in good agreement with
the covariant ChPT analysis therein of lattice QCD results, $\langle
r_A^2 \rangle = 0.31(4)~\textrm{fm}^2$\cite{Alvarado:2026kny}, and
those lattice QCD extractions leading to lower values, albeit in
tension with experimental determinations.

\section{A simple semiempirical representation and final remarks}

In the narrow width approximation one gets the pure meson-dominance
result already discussed in~\cite{Masjuan:2012sk,Amaro:2015lga} 
\begin{eqnarray}
\Omega_{a_1} (-Q^2) \Omega_{a_1'} (-Q^2) \to \frac{m_{a_1}^2}{m_{a_1}^2+Q^2}
\frac{m_{a_1'}^2}{m_{a_1'}^2+Q^2} \, .
\end{eqnarray}
In those works, the half-width rule provided a naive and rather
generous uncertainty estimate, comparable to the available lattice
uncertainties at the moment. For the finite width case, we have
checked that the functional form of the approximation works in the
range $0 \le Q^2 \le 2 {\rm GeV}^2$, but with slightly displayed axial
masses with respect to ones used as a reference in the
construction. Using different models for $\Pi(s)$ fulfilling
analyticity conditions, we have also checked that the induced
uncertainties are much smaller than the half-width rule.

Our calculation involves no fit to data, and after a quantitative
estimate of uncertainties is assessed, can be efficiently summarized by
the semiempirical formula featuring corrections to axial meson
dominance
\begin{multline}
 G_A(-Q^2) \approx  g_A    \frac{m_{a_1}^2}{m_{a_1}^2+Q^2} \frac{m_{a_1'}^2}{m_{a_1'}^2+Q^2} \frac{1}{(1+Q^2/4m_N^2)^{2 \epsilon}} \\
\times \left[ \frac{1+Q^2/4 m_N^2}{1+Q^2/(m_\pi+m_\rho)^2}   \right]^{\overline{\delta}_{\rho\triangleleft}/ \pi}
      \left[ \frac{1+Q^2/4 m_N^2}{1+Q^2/(m_\pi+m_\sigma)^2}   \right]^{\overline{\delta}_{\sigma\triangleleft}/ \pi} 
\end{multline}
in terms of an average triangle-phase value
 $\overline{\delta}_{\rho\triangleleft} \in
[0.2,0.72] $ and  $\overline{\delta}_{\sigma\triangleleft} \in
[0.02,0.08] $  and an asymptotic
radial Regge exponent $\epsilon = 0.1-0.2$, from a matching to pQCD in a
wide range of onset scales. Besides, we note that there is a trend to
cancellation from the triangle and the Regge contributions containing
the nucleon mass, since $\bar \delta/\pi \le 1/2$ and
$\epsilon=0.1-0.2$. This simple approximation clearly shows that the
genuine $3\pi$ cut structure is rather irrelevant as compared to the
$\pi \rho$ cut one.

We hope this observation to be a useful guideline for both
experimental as well as lattice QCD data analyses such as the
popular z-expansion~\cite{Bhattacharya:2011ah} which aims at a model
independent extraction of $G_A(-Q^2)$. This representation has become 
lately the {\it primary} source of information of
data sets and their uncertainties.  Let us remind that, by design, the
z-expansion has an asymptotic expansion $ G_A (-Q^2 ) = a_0 Q^{-4} +
a_1 Q^{-5} + a_2 Q^{-6} + \dots $. The leading behaviour is in
contrast with the rigorous pQCD, $ 1/Q^4 (\log Q^2)^2 $ and also at
odds with our superconvergence sum rules preserving Regge behaviour
$Q^{-4-2 \epsilon}$. The odd-power subleading terms, $Q^{-5}$ and
further are a direct consequence of the built-in $3\pi$ square-root
singularity, $\sqrt{s-9 m_\pi^2}$, quite different from the field
theoretical behaviour obtained from a three-subtracted dispersion
relation, $ (s-9m_\pi^2)^2 \log (s-9 m_\pi^2) $. From this point it
would interesting to implement these physical requirements in the
theoretical analysis.

\section{Conclusions}

The minimal \textit{ansatz} within the axial-vector meson dominance framework, 
with $a_1$ and $a_1'$ masses taken from the PDG, provides the bulk 
of the axial nucleon form factor in the spacelike region 
$0 \le Q^2 \le 2~{\rm GeV}^2$, as it has been
discussed previously. A rule-of-thumb estimate based on the half-width
rule yields generous theoretical error bands, which are in marginal agreement
with the recent lattice QCD analysis, but disagree with
the MINER$\nu$A measurements.

In this work we have analyzed further corrections on top of this scheme,
implementing rigorous superconvergent sum rules from QCD resting on
two gluon exchange at high energies and implementing chiral $3\pi$
corrections in the threshold region. Besides, we have analyzed the
role of $\pi \sigma$ and $\pi \rho$ triangle singularities and the
role of finite widths in the existing $a_1$ and $a_1'$ axial mesons
below $N \bar N$ threshold. We have also modeled the contribution of
an infinite tower of radially excited axial mesons from $\bar N N$ up
to the pQCD onset. Apart from the overall normalization to the physical
axial coupling constant, the energy dependence of our calculation,
including uncertainties, is a genuine prediction independent of 
experimental data or lattice results.

Overall, we find that the main corrections to the simple axial-meson
dominance arise from the $\rho \pi$ and, to a lesser extent, the
$\pi \sigma$ triangles. The infinite Regge tower of radially
excited meson states, as well as finite-width effects, provide smaller
corrections. By contrast, both ChPT and pQCD effects are negligible 
compared with the overall uncertainty of the calculation. We confirm, with a better
confidence level, the agreement with lattice QCD and disagreement with
MINER$\nu$A.
In a similar context, it will be interesting to analyze the impact 
of triangles for the pseudoscalar form factor, and the Goldberger--Treiman 
discrepancy, that we studied in Ref.~\cite{RuizArriola:2025omi}.

{\sl This work is supported by MICIU (Spain) under grant
No. PID2023.147072NB.I00 and Junta de Andaluc\'\i a FQM225.} 

\appendix

\section{Input parameters}\label{app:LoopAndInputs}

In the loop calculation, the Dirac coupling has been modeled with the 
help of chiral Lagrangians to ensure transversity by adopting 
an external source $-g_{\rho NN}\vec{\rho}_\mu \bar q\gamma_\mu\frac{\vec\sigma}{2}q$.
This imply the additional last diagrams in Fig.~\ref{fig:RpiLHC}.   
Additionally, the following phenomenological interactions have been used
   \begin{eqnarray}
 &&     -f_{\rho NN}\bar N\frac{\sigma^{\mu\nu}}{4m_N}\frac{\vec\tau}{2}N\vec{\rho}_{\mu\nu}, \,
      g_{\sigma N N}\bar NN\sigma, \, \\ 
&&      g_{a\rho\pi} \vec{a}^\mu (\vec{\rho}_\mu\!\wedge\!\vec{\pi}), \,
      -g_{a\sigma\pi}\vec{a}_\mu(\overset{\leftrightarrow}{\vec{\pi}\partial^\mu}\sigma) .
   \end{eqnarray}
We note that, for the Pauli coupling, the calculation also leads to 
an additional $-B_0$ term. This however lacks a left-hand cut and is 
equivalent to a point-like coupling that can be reabsorbed in the 
unknown subtraction constant.  
The Dirac and Pauli couplings can be estimated modeling the corresponding 
isovector vector form factors~\cite{Masjuan:2012sk} with two and three 
resonances, respectively, that leads to
   \begin{align}
      f_{\rho NN} &= g_{\rho NN}(\kappa_p -\kappa_n)/(1-m_\rho^2 / m_{\rho''}^2) \simeq 4.5 g_{\rho NN} , \nonumber \\ 
      g_{\rho NN} &= (m_\rho/F_\rho)/(1-m_\rho^2 / m_{\rho'}^2) \simeq 7 .
   \end{align}
The decay constants are defined as
   \begin{align}
      \bra{0} \bar q\gamma^\mu(\gamma^5)\frac{\tau^i}{2}q  \ket{V^j(A^j)} = \delta^{ij} F_{V(A)}m_{V(A)}\epsilon^\mu.
   \end{align}
and we adopt $F_\rho=157$~MeV $F_{a_1}=1.31(25) F_\pi$~\cite{Friot:2004ba}.
For the $\sigma$, we take $g_{\sigma NN} =12$ from Ref.~\cite{CalleCordon:2009pit}.

Finally, the $g_{a_1R\pi}$ couplings can be obtained from the 
corresponding decay widths
\begin{eqnarray}
   g_{a_1\rho\pi}^2 &=& \frac{8\pi m_{a_1} \Gamma_{a_1} \textrm{BR}_{\rho\pi}}{ \frac{\lambda_{\pi\rho}^{1/2}}{m_{a_1}^2}\left(1 +\frac{\lambda_{\pi\rho}}{12m_\rho^2 m_{a_1}^2} \right) }, \\
   g_{a_1\sigma\pi}^2 &=& \frac{48\pi \Gamma_{a_1}\textrm{BR}_{\sigma\pi}}{m_{a_1}} \left( \frac{m_{a_1}^2}{\lambda_{\pi S}^{1/2}} \right)^3 \, ,
\end{eqnarray}
leading to $g_{a_1\rho\pi} = 4.4(5)~\textrm{GeV}$ (assuming $\textrm{BR}_{\rho\pi}\simeq 100\%$) and $g_{a_1\sigma\pi} = 7.6(8)\sqrt{\textrm{BR}_{\sigma\pi}}$. 
The latter BR is not precisely known experimentally and we adopt the theoretical result $\textrm{BR}_{\sigma\pi}=9\%$ from Ref.~\cite{Molina:2021awn}.
The sign of the coupling constants has been chosen to lead to positive 
spectral functions, as predicted by ChPT at low energies.

\bibliography{dispersive-axial}

\end{document}